\documentclass{sig-alternate}
\usepackage{booktabs}
\usepackage{hyperref}  
\usepackage{color}
\usepackage{cleveref}

  \newif\ifdraft
  \draftfalse

\newcommand{\andscomment}[1]{\ifdraft{\color{red} \textbf{[AE]: {#1}}}\else{\vspace{0ex}}\fi}
\newcommand{\giascomment}[1]{\ifdraft{\color{green} \textbf{[GB]: {#1}}}\else{\vspace{0ex}}\fi}

\permission{Publication rights licensed to ACM. ACM acknowledges that this contribution was authored or co-authored by an employee, contractor or affiliate of a national government. As such, the Government retains a nonexclusive, royalty-free right to publish or reproduce this article, or to allow others to do so, for Government purposes only.}
\conferenceinfo{SAC 2015}{April 13-17, 2015, Salamanca, Spain.\\
Copyright is held by the owner/author(s). Publication rights 
licensed to ACM.}
\copyrightetc{ACM \the\acmcopyr}
\crdata{978-1-4503-3196-8/15/04 \$15.00.\\
\url{http://dx.doi.org/10.1145/2695664.2695761}}

\title{On the Impact of Entity Linking\\ in Microblog Real-Time Filtering}

\numberofauthors{4}
\author{ Giacomo Berardi, Diego Ceccarelli, Andrea Esuli and Diego Marcheggiani \\
\affaddr{Istituto di Scienza e Tecnologie dell'Informazione}\\
\affaddr{Consiglio Nazionale delle Ricerche}\\
\affaddr{via Giuseppe Moruzzi, 1, 56124 Pisa, Italy}\\
\email{firstname.lastname@isti.cnr.it}}

\begin{document}

\maketitle

\begin{abstract}
Microblogging is a 
model of content sharing 
in which
the
temporal locality of posts with respect to important events, either of foreseeable or unforeseeable nature, makes applications of real-time filtering 
of great practical interest.

We
propose the use of Entity Linking (EL) in order to improve the retrieval effectiveness, by enriching the representation of microblog posts and filtering queries.
EL is the process of recognizing in an unstructured text the mention of relevant entities described in a knowledge base. 
EL of short pieces of text is a difficult task,
but
it is also a scenario in which the information EL adds to the text can have a substantial impact on the retrieval process.

We implement a start-of-the-art filtering method, based on the best systems from the
TREC Microblog track real-time adhoc retrieval and filtering tasks
, and extend it with a Wikipedia-based EL method.
Results 
show that the use of EL significantly improves over non-EL based versions of the filtering methods.


\end{abstract}





\section{Introduction}
\label{sec:intro}
\noindent Microblogging has gained great popularity in the last years, with Twitter
leading the field with about $500$M of tweets per day and $255$M of monthly active users\footnote{\url{http://goo.gl/46oMx9} \url{http://goo.gl/J05X3D}}.
Novice users can be overwhelmed by such a sheer amount of information, and even technically skilled users can have a hard time when searching for some specific information, especially about recent or currently happening events.
For these reasons the real-time filtering problem recently emerged as a relevant IR problem, as shown by the TREC Microblog Track
\cite{microblegtrec2012}.

The task of microblog real-time filtering starts with just a query and then a time-sorted stream of microposts (tweets) must be processed filtering out tweets that are non-relevant with respect to the query.
The real-time aspect of the process characterizes the problem by (i) imposing limits on the computational cost of the proposed solutions, and (ii) generating a stream of relevance feedback on the tweets that are marked as relevant.
The filtering component of the search system must be able to quickly take the filtering decision on each micropost, avoiding to be a bottleneck in the presentation of search results to the user.
The filtering process starts with very little relevance information; the feedback that accumulates during the stream processing is a crucial information that the filtering method can exploit, if it can do it efficiently, to tune its parameters on the fly.

In this paper we propose and investigate the use of Entity Linking (EL) to enrich microposts and queries representations in order to improve the 
filtering process.
The goal of EL is to recognize in an unstructured text the mention of relevant entities described in a knowledge base.
Wikipedia\footnote{\url{http://wikipedia.org}} 
is usually adopted as the knowledge base, with each of its pages denoting an entity.
For example, the mentions of ``TREC'' in this paper can be linked to the entity denoted by the ``Text Retrieval Conference'' page\footnote{\url{http://en.wikipedia.org/wiki/Text\_Retrieval\_Conference}} in Wikipedia.

The purpose of using EL in microblog retrieval is mainly toward improving recall, as EL can link together the various ways an entity can be mentioned.
For example, EL enables to connect the query ``Michael Schumacher health conditions'', to a relevant piece of text in which the entity is mentioned using a different expression ``Yes! Schumi is out of the coma!''.
In short pieces of text such as microposts it is likely to expect a preference for shorter versions of names, just to fit within space limitations, or for names that are more pertinent with respect to the topic of the post, e.g., preferring ``French President'' or ``Fran\c{c}ois Hollande'' when either referring to the public role or to the personal life events, but no strong assumptions can be made, specially in the case of Twitter in which almost every relevant tweet has a distinct author.

A secondary purpose of EL in microblog retrieval can be toward improving precision, as EL can help to solve ambiguities in text.
However, EL methods usually rely on the presence of mentions of many different entities in the piece of text being analyzed in order to resolve the ambiguities that some of the mentions may present, and short texts such as microposts may often contain an insufficient number of mentions to allow the method to solve the ambiguities.
Moreover, in order to perform a complete disambiguation, some EL methods require a relatively costly graph-based computation for each processed text, which makes impossible to use them in a real-time setup.

We implement a state-of-the-art filtering method, based on the best systems from the TREC Microblog real-time filtering track, and extend it with a Wikipedia-based EL method.
Results of comparative experiments show that EL-aided filtering methods obtain a significantly better performance over non-EL methods.
Our solutions are effective but also efficient,
e.g., they can run on a personal device that gets an input stream, via API, from a microblog infrastructure.

\section{Problem Description}\label{sec:filter}
The task of real-time filtering, as described in the TREC competition, consists in filtering a time-sorted stream of tweets $S=\{s_1,s_2, ..., s_n\}$, by classifying each one as either relevant or non-relevant with respect to a given query $q$.
In the case a tweet $s_t\in S$ is classified as relevant for the query, it is possible to get a relevance feedback stating the correct relevance judgment of $s_t$.
The relevance feedback can be used to update the filtering model for the classification of tweets that have been posted after the time $t$.

\subsection{A Supervised Filtering Approach}\label{sec:supervised_f}
Following the current state-of-the-art system for the task of real-time filtering \cite{AlbakourMO13}, we adopt a supervised approach well-known for its effectiveness in adapting filtering tasks (e.g., \cite{zhang01thebias}), namely Incremental Rocchio classifier \cite{Allan96}.
Incremental Rocchio allows to create a user profile using both relevant and not relevant documents, tweets in our case. 
Each time a new tweet is proposed to the user, Incremental Rocchio checks whether the new tweet is relevant according to the user profile.

Formally, the user profile is calculated as a centroid $\vec{c}_t$ in the vector space:
\begin{equation}\label{eq:Rocchio}
\vec{c}_{t} = \frac{\alpha}{|R_t|} \cdot \sum_{s_i \in R_t} \vec{s}_i - \frac{\beta}{|N_t|} \cdot \sum_{s_i \in N_t} \vec{s}_i
\end{equation}
Where $R_t$ is the set of relevant tweets at time $t$, $N_t$ is the set of non-relevant tweets a time $t$, $\alpha$ and $\beta$ are parameters that weight the ``importance'' of relevant and non-relevant tweets respectively. 
$\vec{s}_i$ represents the vectorial transformation of the tweet $s_i$. 
The index $t$ indicates the time-step in the streaming $S$; given the possibility of a relevance feedback, the centroid $\vec{c}_{t}$ can change through time.

In order to check if a new tweet $s_t$ is relevant for the query, the cosine distance between the centroid $\vec{c}_t$ and the vectorial representation of the tweet $\vec{s}_t$ is calculated.
If the distance is lower than a certain threshold $\eta$ (determined on a validation set, see \Cref{sec:experiments}) the tweet is classified as relevant, otherwise as non relevant. 
In the case in which the tweet is classified relevant a relevance feedback is obtained.
If the tweet is actually relevant it will be added to the set $R_{t+1}$ otherwise it will be added to the set $N_{t+1}$.
The centroid $\vec{c}_{t+1}$ is then updated to take into account the new information.

Following the guidelines of the microblog track, at the beginning of the filtering process $R_{0}$ is composed by the query and the first relevant tweet.

\subsubsection{Features}
In our reference system, which does not use EL, we convert tweets to their vectorial representation by extracting a variety of features.
We take as a feature each word (stopwords are removed), and also its stemmed version, obtained from the Lancaster stemmer \cite{Paice90Another}, which resulted to be more effective than other stemmers.
Stems are marked as distinct features from words that have the same spelling, e.g., ``run'' and ``stem:run'', in order to avoid unwanted alterations of word frequencies.
We take word bigrams by using a sliding window on text; stopwords are not removed in this case.
The title of Web pages pointed by the URLs appearing in tweets may contain information that is useful to determine their context and thus their relevance; we retrieve titles from linked Web pages and we extract the same features as above.
Hashtags may be composed of words that are useful to determine relevance; we perform Viterbi algorithm-based hashtag segmentation \cite{Berardi2011}, e.g., ``\#royalvisitusa'' becomes ``royal visit usa''.
Hashtags are added as features in both original and segmented forms.
We use $tf\cdot idf$ weighting, which, for our filtering method, resulted the best performer in experiments on the validation set comparing various weighting models proposed for the track.

\subsubsection{Filtering}
Han et al.\ \cite{han2012hit} observed that the presence of a URL, which may either point to a Web page or an image, in a tweet is a relevant hint for the relevance of a tweet.
For example, 84\% of the set of relevant tweets in the validation set contain at least a URL, while this value is only 23\% for non-relevant ones.
In \cite{han2012hit} a large boost in precision, with a symmetric loss in recall, has been reported by considering as relevant only tweets that contain at least a URL (the ``hitUWT'' run).
We tested two versions of the reference system, one that only uses the Incremental Rocchio classifier (\textsf{IncRoc} in \Cref{tab:results}), and one that marks as non-relevant any tweet that does not contain a URL (\textsf{IncRocU} in \Cref{tab:results})

For the binary classification by the Incremental Rocchio classifier, the centroid is initially determined on the query vector $\vec{q}$ and the vector of the first relevant tweet of the stream, which is given for each query of the dataset.
The \emph{idf} weights are initially computed on the last 1000 tweets before the first relevant tweet.
The costs of the centroid update and the cosine similarity is linear with the size
of the centroid (i.e., the size of the set of features appearing in relevant tweets), when using efficient sparse data structures and keeping a dictionary of the \emph{idf} weights of the features.

\section{Entity Linking} \label{sec:el}
EL allows to connect small text fragments in a document with entities contained in a given knowledge base, e.g., Wikipedia. 
Given a plain document the linking is usually performed in 2 steps: 
the text fragments that could refer to an entity (called \emph{spots}, \emph{mentions}, or \emph{surface forms}) are identified. 
Since a mention may represent several different \emph{entities} (e.g., the mention ``Italy'' can refer to the country or the Italian football team), a \emph{disambiguation} step is performed, where the correct entity is selected among the candidates.

Let us define $M=\{m_{1}, m_{2}, \dots, m_{|M|}\}$ and $E=\{e_{1},$ $e_{2}, \dots, e_{|E|}\}$ as the sets containing respectively all the mentions and the entities (articles) of Wikipedia. 
Each mention links to one or more entities. 
In this work we rely on two important measures used for EL: the \emph{link probability} and the \emph{commonness}.
Given a mention $m$, $lp(m)$, the link probability represents the probability of the text $m$ to be a link to an entity in Wikipedia.
This property allows us to discriminate mentions that link, with a high probability, to some entity from those referring to an entity only occasionally. 
For example, the mention ``the'' occurs a huge number of times in Wikipedia, but only in a few cases links to the Wikipedia page about the English articles.
Given an entity $e$ and a mention $m$,  $cm(m, e)$ represents the commonness, i.e., the probability $p(m|e)$ that a mention $m$ links to $e$.
For each mention, the sum of the commonness of all its related entities is $1$.
The commonness determines the strength of the relation mention-entity.
These values are computed on the Wikipedia collection used by the entity linker.

\section{Entity Linking-Aided Real-Time Filtering}\label{sec:elfilter}

As already discussed, the main purpose of using EL in filtering is to increase the possibility of linking semantically related information expressed in different forms.

The text of queries, tweets, and the title of Web pages linked by tweets, are analyzed by an entity linker that produces a list of found mentions, each one paired, with a link probability value, with a \emph{candidate entity}.
We also define as the \emph{surface forms set} ($SF$) of an entity $e$ the set of mentions $SF(e) = \{m \;| \; cm(e,m) > 0 \} $, i.e., all the mentions that link to the entity $e$ at least once. 
For example, the surface form set for the entity \emph{Diego Armando Maradona} (DAM in the following), is constituted by the mentions ``argentine legend'', ``diego armando maradona'', ``el diego'', etc.

It is worth noting that in our efficiency- and recall-oriented approach, we do not perform disambiguation based on the relations among the candidate entities in the knowledge base graph.
Our approach mainly relies on commonness, and is inspired by the simple method evaluated by Meij et al.\ \cite{meij2012adding}. 
In that paper, Meij et al.\ prove that linking the entity with highest commonness with respect to the mention has a good performance on tweets. 
In particular they show that the commonness method applied on tweets outperforms the traditional state-of-the-art methods~\cite{Ferragina:2010,mendes2011dbpedia,milne2008learning}.
The authors of \cite{meij2012adding} also propose a machine learning approach to improve the performance of EL on tweets, but we decided to rely on \emph{commonness} for its simplicity and also because it represents a good trade-off between quality and efficiency.

We explore three ways of expanding the features space in which tweets are represented by means of EL.
Two of the methods leverage on the bipartite graph of mentions and entities.
Navigating that graph is a critical process, since it is necessary to find a sweet point in the exploration range, in order to be able to find new relevant information without adding undesired noise.

\begin{figure}[h]
\centering
\includegraphics[width=0.8\columnwidth]{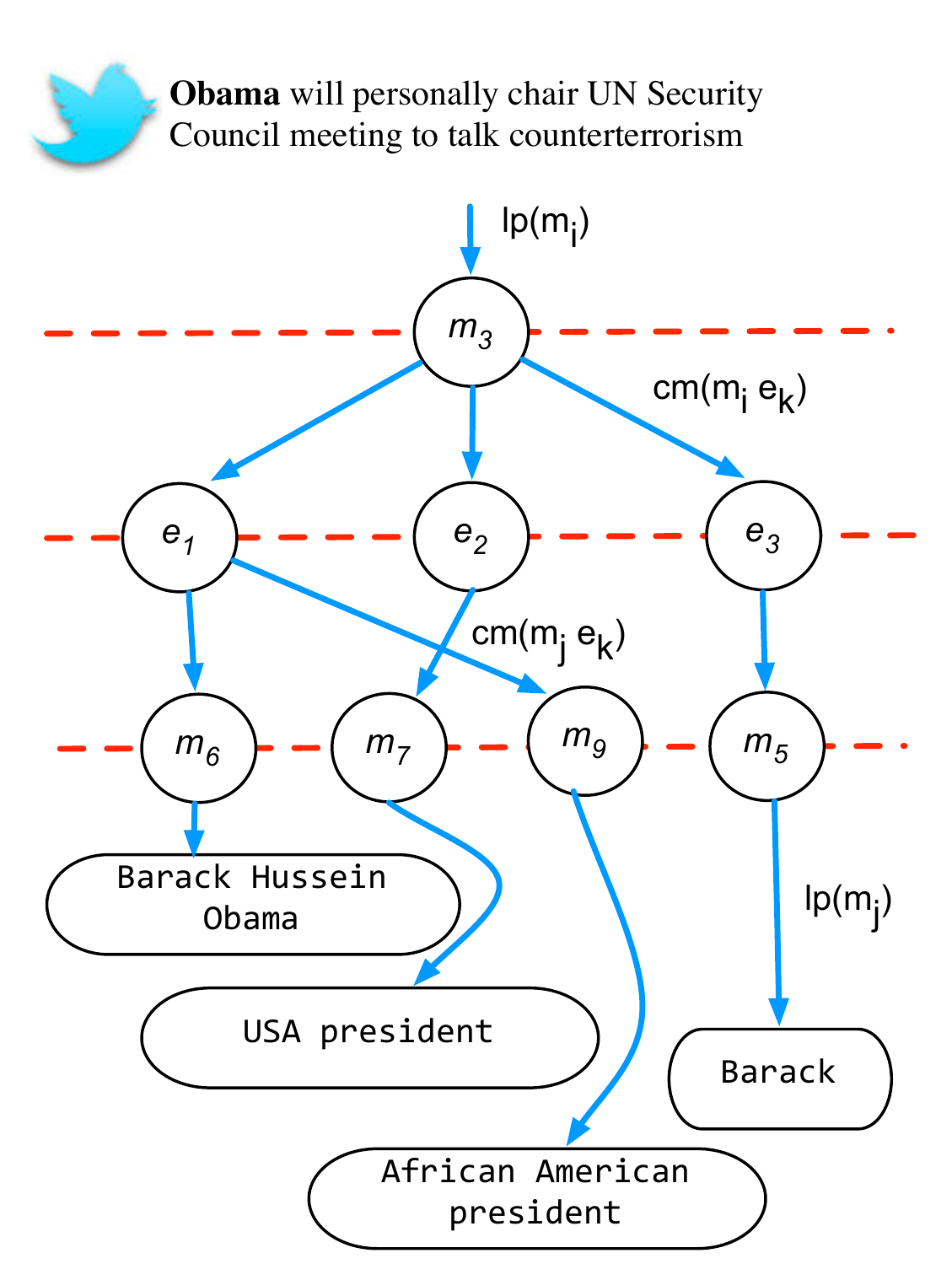}
\caption{The directed acyclic graph used when computing feature expansion. 
In this example, ``Obama'' matches a mention $m_{3}$ that refers to several entities, each one linking to a number of other mentions.
For each of these mentions the relative surface form is shown.
The original mention and the surface forms found navigating the graph are connected through paths, in which edges are weighted by the $lp$ and $cm$ functions, as indicated.}
\label{fig:graph_el}
\end{figure}

\subsection{Features expansion by highlighting mentions} \label{sec:ment_exp}

This method (dubbed \textsf{Exp1} hereafter) adds a new feature for each mention found in text.
For example, for ``el diego'', it adds the feature ``ment:el\_diego''.
The rationale behind this method is to promote, by replicating it, the weight of mention-related components of text in the determination of similarities between vectors.
A threshold for the minimum link probability for which a piece of text can be considered a relevant mention must be determined experimentally (see \Cref{sec:settings}).
Note that different surface forms for the same entity are not linked by this simple method, an issue addressed by the other two methods.

\subsection{Features expansion by exploring entities} \label{sec:ent_exp}

In Figure \ref{fig:graph_el} we represent an example of a graph generated with the information obtained by the entity linker.
The nodes of the graph are mentions and entities, the edges are weighted by the probabilities given by $lp$ and $cm$.

The second expansion method (dubbed \textsf{Exp2}) selects the most relevant entities returned by the entity linker for all the mentions in a tweet, and creates new features from them.
We define $p_{en}(m_{i}, e_{k})$, the probability of linking the mention $m_{i}$
to the entity $e_{k}$, as:
\begin{equation}\label{eq:ent_exp}
    p_{en}(m_{i}, e_{k}) = lp(m_{i}) \cdot cm(m_{i}, e_{k})
\end{equation}
Candidates entities are ranked by $p_{en}$ and those with a score higher than a parameter $\rho$, optimized on a validation set, are selected.
With \textsf{Exp2} the mentions ``el diego'' and ``Armando Maradona'' are expanded into the same feature ``ent:eDAM'', i.e., the unique identifier of the entity, thus effectively enabling the possibility of linking two pieces of text in which the same entities are mentioned using different mentions.

\subsection{Features expansion by exploring surface forms} \label{sec:alias_exp}

The third expansion method (dubbed \textsf{Exp3}) expands a tweet with all the alternative surface forms of the entities in the tweets.
We define $p_{sf}(m_{i}, e_{k}, m_{j})$ as the probability of associating the surface form $m_{j}$ to the mention $m_{i}$ through the entity $e_{k}$.
\begin{equation}\label{eq:alias_exp}
    \begin{split}
        p_{sf}(m_{i}, e_{k}, m_{j}) = &lp(m_{i}) \cdot cm(m_{i}, e_{k}) \\
                                      \cdot &lp(m_{j}) \cdot cm(m_{j}, e_{k})
    \end{split}
\end{equation}
where $e_{k}$ is an entity which connects the mentions $m_{i}$ and $m_{j}$.
Similarly to \textsf{Exp2}, a parameter $\rho$ sets the threshold for the selection of which candidates have to be expanded into new features.

In the case of \textsf{Exp3} the example mention ``el diego'' is thus expanded into a set of features of all the surface forms $m_{j}$ for the entity \textbf{DAM} for which $p_{sf}($``el diego''$,DAM,m_{j})>\rho$, e.g., ``sf:armando\_maradona'', ``sf:argentine\_legend'', \emph{etc}.

With \textsf{Exp3}, the entities with more surface forms are likely to have a higher weight in the determination of similarity between vectors, due to the larger number of features generated.
Depending on the $lp(m_{i})$ and $cm(m_{i}, e_{k})$ values for the original mention $m_{i}$ from which $p_{sf}$ is computed, the surface forms for an entity $e_{k}$ gets different weights with respect to other initial mentions $m_{x}$. 
This results in different sets of features being added to vectors depending on the initial mention, thus adding a grading in the similarity between vectors referring to the same entity using different forms.

Moverover, ambiguous surface forms, such as ``Formula One driver'' may generate common features for pieces of text in which the original mentions were linked to different, but related, entities, e.g., creating a common feature from the entity ``Nico Rosberg''
and the entity ``Lewis Hamilton'';
this can however result in adding features that link unrelated pieces of text. 
The behaviour is somewhat expected from the \textsf{Exp3}, since it is the most aggressive method of the three; experiments will determine if the benefits from the additional linking possibilities are stronger or weaker than the errors introduced by ambiguous surface forms.



\section{Experiments} \label{sec:experiments}

\subsection{Dataset}\label{sec:dataset}
We compared the methods described in \Cref{sec:filter,sec:elfilter} on the dataset originally made available by TREC for the microblog ad-hoc retrieval task \cite{ounis2011overview} and used also for the real-time filtering task \cite{microblegtrec2012}.

Originally consisting of 16M tweets, due to Twitter policies it must be independently downloaded by each research group, resulting in difference as tweets and accounts are deleted.
Our download resulted in 14M tweets retrieved.
Similarly to \cite{han2012hit}, \Cref{tab:results} reports the recall for the trivial acceptor filtering, \textsf{AllRel}, indicating the coverage of our copy of the dataset with respect to the original one.

The microblog track organizers also provided the relevance judgments for 49 queries.
Ten queries form a validation set that can be used for parameter optimization. 
The other 39 queries form the test set on which we evaluate the methods.
\newpage
\subsection{Evaluation Measures}
As evaluation measures we adopt the same four evaluation measures that have been chosen for the real-time filtering task of the microblog track: $precision$, $recall$, $F_{0.5}$ and $T11SU$.
The $F_{0.5}$ function is the $F_{\beta}$ function with $\beta=0.5$, i.e., preferring $precision$ over $recall$.
The $T11SU$ is the \emph{scaled linear utility}, a measure for the adaptive filtering task \cite{Robertson02thetrec}.
Results are averaged across the 39 test queries.
%


\subsection{Experimental Setting} \label{sec:settings}

All the code of the real-time filtering system we implemented, together with the instructions for reproducing the experiments, is released with an open-source license and is available at \url{https://github.com/giacbrd/CipCipPy}.
For the EL functionalities we used Dexter\footnote{\url{http://dexter.isti.cnr.it/}} \cite{ceccarelli13} a versatile open-source framework for EL.

We have tuned the parameters of the system on the validation set, evaluating them with respect to the $F_{0.5}$ measure.
We have optimized the parameters distinctly for each of the tested setup reported in \Cref{tab:results}.
We have run grid search experiments to optimize the $\alpha$ and $\beta$ parameters of \Cref{eq:Rocchio}, along with the $\eta$ threshold for the classifier.
Results indicate that using non-relevant tweets has a negative impact and it is better to consider only relevant tweets (i.e., $\beta = 0, \alpha = 1$, confirming the similar findings of \cite{AlbakourMO13}).
The actual centroid formula we used is thus:
\begin{equation}\label{eq:rocchiopos}
\vec{c}_{t} = \frac{1}{|R_t|} \cdot \sum_{s_i \in R_t} \vec{s}_i
\end{equation}
Another parameter to tune is $\rho$, which is used to exclude candidate features (either entity ids, or surface forms) with low probability; we have found that its optimal value is the same for all the entity expansion approaches, i.e., 0.1.
We also determined that the value of minimum link probability that a mention should have, in order to be linked, is 0.2. 
This value is however not crucial for the efficacy of the expansion, which is instead strictly dependent on $\rho$.

\subsection{Results}

\Cref{tab:results} reports the results of our methods and a number of methods we compare to.
\textsf{Medians2012} are the median values for all the participants to the 2012 TREC Microblog real-time filtering track;
\textsf{Best2012} is the best performer of that track.
The current best system, with respect to $F_{0.5}$, in real-time filtering is \textsf{CurrentBest}, namely \cite{AlbakourMO13}.

\textsf{AllRel} is a trivial acceptor system which marks all tweets as relevants; this allows to assess the coverage of relevant tweets of the our copy of the corpus, which is a good 95.4\%\footnote{For comparison, \cite{han2012hit} reports 91.5\%.}

\textsf{IncRoc} and \textsf{IncRocU} are our implementation of supervised method we describe in \Cref{sec:supervised_f}, inspired to the state-of-the-art systems.
\textsf{IncRocU} differs from \textsf{IncRoc} as it follows the findings of \cite{han2012hit} and it marks any tweet which does not contain a URL as non-relevant.
Both system obtains very good scores, competitive with the state of the art.
\textsf{IncRocU} obtains a higher recall and $F_{0.5}$ with respect to the state-of-the-art systems, and it is also the system which better balances precision and recall.
Since \textsf{IncRocU} obtains the highest $F_{0.5}$ and $T11SU$ scores, it will be our strong reference system on which we test our expansion methods.


Experiments that use the proposed EL-based expansion methods are listed as \textsf{Exp1}, \textsf{Exp2} and \textsf{Exp3}.
%
%
The \textsf{Exp1} method (\Cref{sec:ment_exp}) obtains a relative +2.2\% improvement in precision, with almost no variation in recall ($F_{0.5}$ and $T11SU$ respectively improve by +3.0\% and +3.4\%).
This indicates that just giving more importance to mentions over the rest of the text allows to better differentiate relevant and non-relevant tweets, following the intuition that mentions of entities strongly characterize the content of tweets.
%
%
The \textsf{Exp2} method (\Cref{sec:ent_exp}) is the top performer, and improves on all the measures (Precision +2.9\%, Recall +2.1\%, $F_{0.5}$ +4.6\%,  $T11SU$ +4.2\%).
The significant increase in recall shows the impact of entity-based features in strengthening the similarities among documents which talk about the same concepts, possibly using different expressions.
Precision increases, though marginally, with respect to \textsf{Exp1}, meaning that the potential noise or ambiguity generated by entity-based expansion is null or irrelevant.
%
%
The \textsf{Exp3} method (\Cref{sec:alias_exp}) instead worsens all the results with respect to the base system, indicating that a more aggressive expansion adds detrimental ambiguity and noise into the text.

We also tested a variation of the \textsf{Exp2} expansion method, \textsf{Exp2-1Ent}, which adds for each mention identified by the linker only the candidate entity with the higher commonness.
This variation follows the work of Meij et al.\ \cite{meij2012adding} (see \Cref{sec:elfilter}), and can be considered as a light disambiguation step.
\textsf{Exp2-1Ent} slightly improves over the base system (not on recall), but it is worse than \textsf{Exp2}, indicating that being more inclusive in adding entities is a better strategy.

 \begin{table}[tbp]
 \centering
\begin{tabular}{lcccc}
  \toprule[1.5pt]
{\bf Method} &  {\bf Precision} & {\bf Recall} & $\mathbf{F_{0.5}}$ & {\bf T11SU} \\
  \midrule
 \textsf{Medians2012} & .177 & .334 & .149 & .208\\
  \textsf{Best2012} & \textbf{.622} & .174 & .334 & \textbf{.412}\\
  \textsf{CurrentBest} \cite{AlbakourMO13} & .421 & .337 & .344 & .361\\
\midrule
 \textsf{AllRel} & .000 & .954 & .000 & .000 \\
\midrule
\textsf{IncRoc} \giascomment{temp!} & .329 & .423 & .323 & .284 \\
 \textsf{IncRocU} & .408 & .382 & .372 & .357 \\
\midrule
 \textsf{Exp1} & .417 & .383 & .383 & .369 \\
 \textsf{Exp2} & .420 & \textbf{.390} & \textbf{.389} & .372 \\
 \textsf{Exp3} & .371 & .341 & .342 & .327 \\
 \textsf{Exp2-1Ent} & .409 & .379 & .377 & .367 \\
  \bottomrule[1.5pt]
 \end{tabular}
 \caption{\label{tab:results}Evaluations of the runs.
 In bold the best result for a specific evaluation measure.}
 \end{table}


To sum up, we started from a strong baseline as \textsf{IncRocU}, and by simply using information in the knowledge base to promote mentions to first class features (\textsf{Exp1}) we improved precision.
When using a proper EL-based method to perform expansion by entities (\textsf{Exp2}), thus enabling linking similar concepts expressed in different forms, also recall, and all the other measures, improved.


\section{Related Work}\label{sec:related}

IR in microblogging has been investigated in the works of Efron \cite{efron2011}, Nagmoti et al.\ \cite{Nagmoti2010} and Weng et al.\  \cite{Weng2010}.
Efron overviews various IR tasks in the domain of microblogging.
Weng et al.\  propose an extension of PageRank algorithm to measure the influence of users in microblog platform such as Twitter.
Nagmoti et al.\ propose several ranking strategies for ad-hoc retrieval in the microblog domain.

The first large scale initiative that has focused on microblogging is the 2011 TREC Microblog Track \cite{ounis2011overview}, with the task of ad-hoc real-time search.
The second edition of TREC Microblog Track \cite{microblegtrec2012}
proposed also the real-time filtering task.
Several research groups have participated to the 
real-time filtering task.
Many of them have faced the filtering task adopting supervised learning methods, and binary classification is performed on tweets by employing several techniques. 
Online learning methods, such as the Rocchio algorithm, have been widely used \cite{liang2012pkuicst,limsopatham2012university}, but also batch learning techniques have been adopted \cite{karimi2012searching,liang2012pkuicst}. 
Among the top performing methods there also methods that are mainly based on the retrieval scores produced by IR systems designed for the ad-hoc search task.
In particular, the system that ranked first \cite{han2012hit} adopted a method designed for the ad-hoc retrieval task (based on learning to rank), performing a reference search on the documents preceding the query and judging any new tweet from the stream as relevant whether its retrieval score is not smaller than the scores of the $m$ most relevant tweets from the reference search.
In the same way, the system that ranked second \cite{zhang2012pris} adopted an ad-hoc retrieval system based on learning to rank and, in order to face the filtering task, it implemented an adaptive threshold mechanism.
A notable work that followed the 2012 TREC Microblog Track real-time filtering task is from Albakour et al.\ \cite{AlbakourMO13}, which proposes a Rocchio-based system extended with a query expansion technique for dealing with sparsity.

EL is widely used in text mining, especially for short texts \cite{Ferragina:2010, Hu:2009}.
Some recent works in EL \cite{dalton14entity, Ren:2014, Schuhmacher:2014} adopt modern approaches for expanding textual representations for queries and documents. 
They use the graph structures returned by the EL frameworks to obtain new features that are semantically related to the pieces of text being analyzed.

Use of EL in microblog ad-hoc retrieval has been proposed by Feltoni and Gasparetti \cite{Feltoni2012}.
Given a query, the top 15 results from a standard IR system are given in input to a wikification service; the annotated entities are then compared to the terms in the original query and those with the highest semantic relatedness, as determined by another service that leverages on the Wikipedia link graph, are then used for query expansion on the IR system.
In addition to the inherent differences between the ad-hoc retrieval and real-time filtering tasks, 
our work differs from \cite{Feltoni2012} as we do not use an IR system, we do not use EL output from tweets to expand the query,
and our method automatically adapts its parameters based on relevance feedback.
Our method is also computationally lighter, as it does not perform full wikification of text nor it uses any complex graph-based processing.


\section{Conclusions\andscomment{and Future Work}} \label{sec:conclusion}

Real-time filtering of microblogs is a non-trivial task. 
It poses a combination of challenges which differentiate it from related tasks such as ad-hoc retrieval on microblogs or filtering from other sources of content: short texts, use of jargon/hashtags, real-time content evolution and feedback, compromises between responsiveness and efficacy.

We implemented a reference system that is competitive with (and for some metrics better than) the current state of the art.
On top of that system we tested the impact of different ways of using EL to enrich the content of tweets.
Two of the methods, \textsf{Exp2} over all, considerably improved the non-EL results, showing how even a simple and efficient method for incorporating the information from an external knowledge can have an impact on retrieval applications.
We consider that our intuition of using EL to improve the filtering effectiveness has been confirmed by the results.

Even though the expansion methods were originally designed to improve recall, results have shown a sensible improvement in precision too, indicating that mentions and entities are a key element of content in order to determine its relevance with respect to a query.


\andscomment{Future work will explore the use of more complex use of EL, including full wikification techniques.}

\vspace{0.2cm}
{ \footnotesize
{\bf {Acknowledgements}} This work was partially supported the Regional (Tuscany) project SECURE! (POR CReO
FESR 2007/2011)}

\bibliographystyle{abbrv}


\end{document}